\documentclass[twocolumn,showpacs,aps,dvips]{revtex4}
\usepackage{amsfonts,amssymb,amsmath,graphics}
\usepackage{graphicx,wrapfig}
\newcommand*{\dis}{\displaystyle}
\newcommand*{\bs}{\boldsymbol}

\begin{document}
\title{ Relativistic Nucleus-Nucleus Collisions without Hydrodynamics }

\author{D.V. Anchishkin$^a$, S.N. Yezhov$^{\, b}$}
\affiliation{$^a$Bogolyubov Institute for Theoretical Physics,
             03068 Kiev, Ukraine}
\affiliation{$^b$Taras Shevchenko National University, 03022 Kiev,
             Ukraine }



\begin{abstract}
The partition function of nonequilibrium distribution which we
recently obtained [arXiv:0802.0259] in the framework of the maximum
isotropization model (MIM) is exploited to extract physical information
from experimental data on the proton rapidity and transverse mass
distributions.
We propose to partition all interacting nucleons into ensembles in
accordance with the number of collisions.
We analyze experimental rapidity distribution and get the number of
particles in every collision ensemble.
We argue that even a large number of effective nucleon collisions
cannot lead to thermalization of nucleon system;
the thermal source which describes the proton distribution in central
rapidity
region arises as a result of fast thermalization of the parton degrees
of freedom.
The obtained number of nucleons which corresponds to the thermal
contribution is treated as a ``nucleon power'' of the created quark-gluon
plasma in a particular experiment.

\end{abstract}


\maketitle

The main goal of the investigations of the collisions of relativistic
nuclei
is extraction of a physical information about nuclear matter and its
constituents.
It is a matter of fact that we can get know more about quarks and gluons
(constituents) just under extreme conditions, i.e. at high densities and
temperatures.
During last two decades one of the celebrated tools on the way of the theoretical
investigations of this subject was
relativistic hydrodynamics (RHD) which started to be applied to
elementary particle physics from the famous Landau's paper
\cite{landau-1953}.

Applying RHD one can partially describe experimental data and get
know that the matter created in relativistic nucleus collisions
(RNC) can be regarded on some stage of evolution as a continues
one, i.e. as a liquid. Moreover, as was discovered in BNL, it can
be regarded even as a perfect fluid \cite{Gyulassy-0403032} which
consistent with a description of the created quark gluon plasma
(QGP). The main physical quantities which can be extracted from
experimental data exploiting the RHD approach are the collective
(hydrodynamical) velocity and the elliptic flow parameter $v2$ of
the fireball expansion. Unfortunately, all other physical
information is hidden in sophisticated numerical codes which solve
Euler hydrodynamic equations of motion.

In the present letter we propose approach to description of
relativistic heavy-ion collisions which allows to extract the
physical information from experimental data on the basis of
transparent analytical model.

{\bf Maximum Isotropization Model.}
We separate all amount of registered nucleons into
groups (ensembles) in accordance with a number of collisions, $M$,
which every nucleon from a particular ensemble has went through.
The first collision ensemble is created by the nucleons which take
part
just in one collision only, then $M=1$, the second ensemble is
created by the nucleons which take part in two collisions only,
$M=2$, and so on.
Every ensemble contributes to momentum
single-particle distribution function which can be written as
\begin{equation}
\frac{d^3 N}{ dp^3}
= \! \!
\sum_{M=M_{\rm min}}^{M_{\rm max}} \! \! C(M)\, D_M({\bs p})
+C_{\rm therm} \, D_{\rm therm}({\bs p})
\, ,
\label{total}
\end{equation}
where the coefficients $C(M)$ and $C_{\rm therm}$ reflect the
weight of the partial contribution from every $M$-th collision
ensemble and thermal distribution, respectively, to the
three-dimensional momentum spectrum, $M = 0$ corresponds to
spectator particles which are not taking into account in this
distribution. For the sake of simplicity we consider collision of
the identical nuclei.
The expression for the partial distribution
functions $D_M({\bs p})$ was derived in Ref.
\cite{anchishkin-ezhov-ujp2007}, in the c.m.s. of colliding nuclei
it reads
\begin{eqnarray}
D_M({\bs p})
=&&
\frac 1{2z_M(\beta)} \, e^{-\beta \omega_p
  -\alpha {\bs p}_\bot^2}
\nonumber \\
&& \times \left[ \, e^{ - \alpha(p_z-k_{0z})^2  }
  + e^{ - \alpha(p_z+k_{0z})^2} \, \right]
\, ,
\label{distrib-D}
\end{eqnarray}
where in the Cartesian coordinate system $\alpha \equiv 3/{2M p_{\rm max}^2 }$,
and in the spheric system  $\alpha \equiv 5/{2M p_{\rm max}^2 }$,
\begin{equation}
z_M(\beta)
= \int d^3k \, e^{ -\beta \omega_k
 - \alpha \left[ {\bs k}_\bot^2+(k_z-k_{0z})^2 \right]  }
 \, ,
\label{partition-func}
\end{equation}
$\beta=1/T$ is the inverse temperature,
$\omega_p=\sqrt{m^2+{\bs p}^2}$, ${\bs p}_\bot= (p_x,p_y)$ and $0z$
is the collision axis.
It is understood from (\ref{distrib-D}) and (\ref{partition-func})
that the quantity $z_M(\beta)$ plays the role of the canonical
single-particle partition function of the $M$-th collision ensemble.
In some sense particular collision ensemble $M$ can be
regarded as a many-particle system frozen at some stage of evolution
on the way to thermal equilibrium ($M \propto$ time).

The thermal distribution reads,
$D_{\rm therm}({\bs p})
=\exp{ [-\beta \omega_p]}/z_{\rm therm}(\beta)$
with $z_{\rm therm}(\beta) = \int d^3k \, e^{ -\beta \omega_k}$.
Note, we separate in (\ref{total}) the thermal contribution due to
its specific role.
It would seem the contribution $D_{\rm therm}({\bs p})$ should appear
in (\ref{total}) as a term of the expansion with respect to partial
contributions, $D_M({\bs p})$,
when the number of collisions is big enough, i.e. when
$M_{\rm max} \to \infty$.
Meanwhile, because of finite life time of the system (fireball)
and hence finite number of elastic and inelastic collisions of
nucleons this limit regime of hadron dynamics, $M\to \infty$, is not
achieved and $M_{\rm max}$ is finite.
However, we include in (\ref{distrib-D}) a thermal source
because it has another nature.
We will return later to the discussion of this matter.

There are two additional quantities in (\ref{distrib-D})-(\ref{partition-func}), $k_{0z}$ and $p_{\rm max}$, which
are the external parameters determined by the particular
experimental conditions.
The values $\pm k_{0z}$ are the initial momenta of nucleons in c.m.s.
Indeed, due to the specifics of heavy-ion collisions we know exactly
the initial momenta of the nucleons in both colliding nuclei.
Two Gaussians in the brackets on the r.h.s. of
(\ref{distrib-D}) reflect a smearing around initial momenta which is
due to collisions of nucleons and were obtained
with the help of the saddle-point approximation. Note, for
$M=1,2,3$ this approximation is not used.

Under the notion ``collision'' we mean elastic rescattering as
well as inelastic scattering (reactions), which include a creation
of secondary particles.
In the transverse direction both nuclei have the same zero initial
momentum.
Then, for both nuclei the smearing around this value is appeared
in (\ref{distrib-D}) as the common factor,
$ \exp{\left[- \alpha {\bs p}_\bot^2 \right]}$.
The covariance of the Gaussian depends on the number of collisions $M$
and the maximally allowed transferred momentum.

The rapidity distribution was obtained after integration of
(\ref{total}) with respect to the nucleon transverse
mass, $m_\bot=(m_N^2+{\bs p}_\bot^2)^{1/2}$, where $m_N$ is the
nucleon mass and rapidity, $y$, is defined as
$\tanh{y}=p_z/\omega_p$.
With respect to new variables one gets,
$d^3p=d\phi \, \omega_p \, m_\bot \, dm_\bot \, dy$.

As a first step of our approach we consider a central
collision of identical nuclei and we assume an azimuth
symmetry of radiation of the particles.
Rapidity spectrum of registered particles looks like
\begin{equation}
\frac{dN}{dy}
=
\sum_{M=M_{\rm min}}^{M_{\rm max}}C(M) \, \varphi_M(y)
 +C_{\rm therm} \, \varphi_{\rm therm}(y) \, ,
\label{y-fit}
\end{equation}
where
\begin{equation}
\varphi_M(y)
= 2\pi \cosh{y} \int_{m_{{\scriptscriptstyle N}}}^\infty  dm_{\perp} \, m_{\perp}^2 \,
 D_M(m_\bot, y)
\, .
\label{phiM}
\end{equation}
To obtain $\varphi_{\rm therm}(y)$ we put $D_{\rm therm}$ in place
of $D_M$ on the r.h.s. of (\ref{phiM}).
Double differential spectrum which depends on the transverse mass
is obtained from (\ref{total})
\begin{eqnarray}
\frac{d^{\,2} N}{ m_\bot dm_\bot dy}
=&& \!\! 2\pi\,
m_\bot \cosh y \Big[\sum_M  C(M) D_M(m_\bot,y)
\nonumber \\
&& \! + \, C_{\rm therm} \, D_{\rm therm}(m_\bot,y) \Big]
\, .
\label{tr-total}
\end{eqnarray}
Usually the mode $M=1$ does not give contribution
to the particular experimental rapidity window.
In this case we can set $C(1)\simeq 0$  and start summation in
(\ref{y-fit}) from $M_{\rm min}=2$.


{\bf Extraction of the physical information from experimental
          data. }
With making use of the rapidity distribution (\ref{y-fit}) we
fit the experimental data on the rapidity distribution
of net protons which were measured at the CERN SPS (NA49
Collaboration) \cite{NA49-PRL-v82p2471-1999}.
The slope parameter $\beta$ was first extracted from double
differential
yield for protons with the use of the thermal distribution.
The proton data is remarkable in that sense that we know exactly the
initial momentum, $k_{0z}$, of every nucleon.
The fit was carried out with a help of the program MINUIT,
variable parameters are coefficients $C(M)$ and $C_{\rm therm}$ and
parameter, which confines the momentum space, $p_{\rm max}$.
The values of the obtained parameters for $T=  1/\beta = 248$~MeV are
shown in Table 1.
All evaluations are carried out in the c.m.s. of colliding nuclei with
use of the spheric coordinate system.

\noindent\begin{tabular}{c c c c c c c  } &&&&&& Table 1 \\
 \hline%
   \multicolumn{1}{|c}{$C(2)$} &   \multicolumn{1}{|c}{$C(3)$} & \multicolumn{1}{|c}{$C(4)$}
 & \multicolumn{1}{|c}{$C(5)$} & \multicolumn{1}{|c}{$C(6)$}
 & \multicolumn{1}{|c}{$C(7)$} & \multicolumn{1}{|c|}{$C(8)$}
\\%
\hline%
   \multicolumn{1}{|c}{4.47} & \multicolumn{1}{|c}{11.9}
 & \multicolumn{1}{|c} {28.9} & \multicolumn{1}{|c} {11.7}
 & \multicolumn{1}{|c} {10.5} & \multicolumn{1}{|c}{9.5}
 & \multicolumn{1}{|c|}{9.57} \\%
 \hline
\end{tabular}

\noindent\begin{tabular}{ c c c c c c c }
 \hline%
  \multicolumn{1}{|c|}{$C(9)$} &   \multicolumn{1}{|c}{$C(10)$}
 & \multicolumn{1}{|c}{$C(11)$} & \multicolumn{1}{|c}{$C(12)$}
 & \multicolumn{1}{|c}{$C_{\rm therm}$}
 & \multicolumn{1}{|c|}{$p_{\rm max}$ (GeV/c)} \\%
\hline%
   \multicolumn{1}{|c}{10.0} & \multicolumn{1}{|c}{10.6}
 & \multicolumn{1}{|c}{11.2}               & \multicolumn{1}{|c}{11.8}
 & \multicolumn{1}{|c}{18.04} & \multicolumn{1}{|c|}{1.275} \\%
 \hline
\end{tabular}

\begin{figure}
\includegraphics[width=0.47\textwidth] {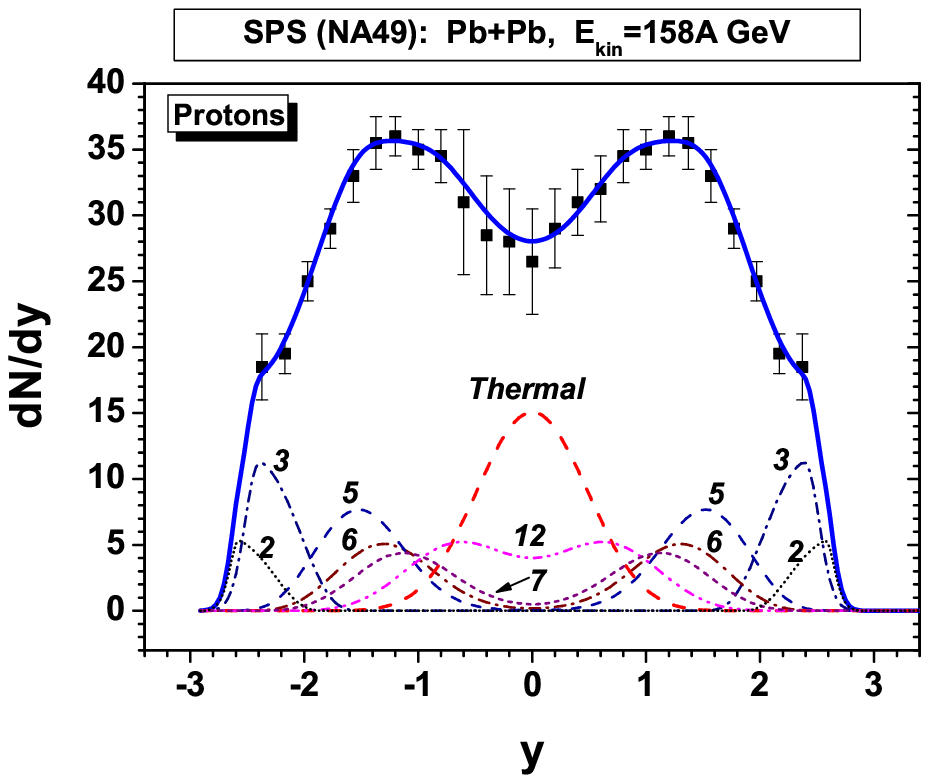}
\includegraphics[width=0.46\textwidth] {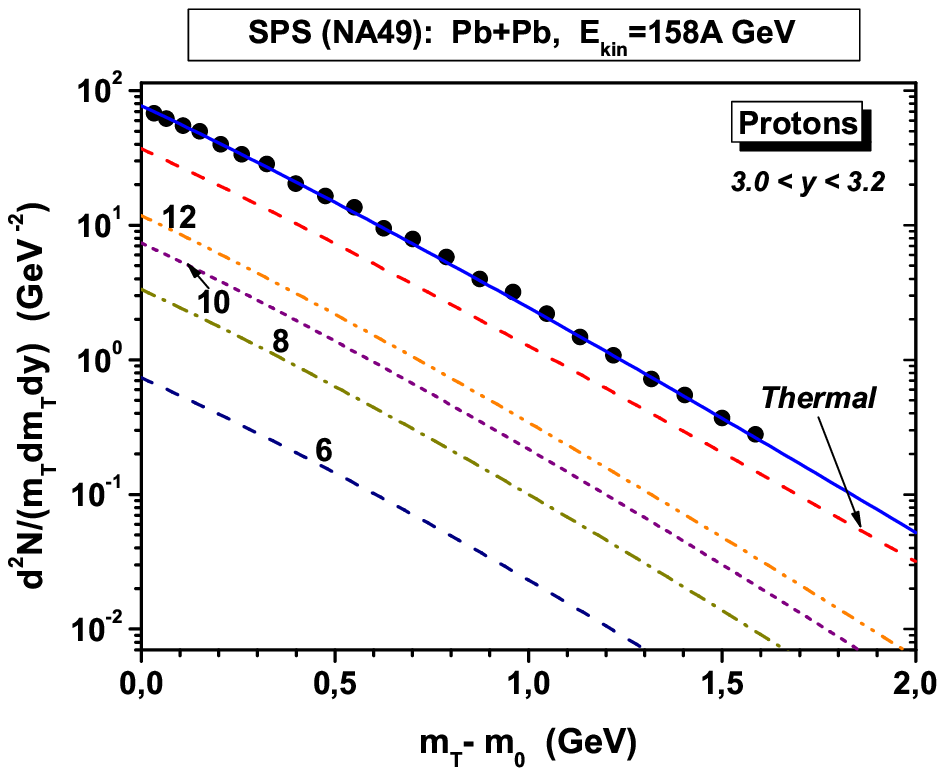}
\caption{
  \noindent
  {\it Upper panel:} The result of the fit (thick solid curve, blue in
  on-line presentation) to experimental
  data \cite{NA49-PRL-v82p2471-1999} on rapidity distribution
  (see Eq. (\ref{y-fit})).
  Broken curves marked by the numbers $M$
  represent the partial  contributions from every collision ensemble,
  $C(M)\cdot \varphi_M(y)$, the thermal contribution is represented by
  central Gaussian-like dashed curve (red in on-line presentation).\\
  {\it Lower panel:}
  The thick solid curve (blue in on-line presentation)
  represents the evaluation of the $m_{\perp}$-spectrum obtained in
  accordance with Eq. (\ref{tr-total})  were we use the same values of
  the coefficients $C(M)$
  which were obtained as a result of the fit of the $dN/dy$ data.
  Experimental data points are from \cite{NA49-PRL-v82p2471-1999}.
  Broken curves marked by the numbers $M$ represent the partial
  contributions from every collision ensemble,
  $2\pi\,m_\bot \cosh y \cdot C(M) D_M(m_\bot,y)$.}
\label{figure:NA49-Mmax12}
\end{figure}

\medskip
\noindent
The obtained theoretical curves together with experimental data are depicted in
Fig.~\ref{figure:NA49-Mmax12}.
Broken curves (see upper panel) marked by the numbers $M$ and  solid
thick curve  (blue in on-line presentation) represent the partial
contributions from every ensemble, $C(M) \cdot \varphi_M(y)$,  and
complete theoretical proton rapidity distribution,
respectively.
The thermal contribution is represented by central
bell-like dashed curve (red in on-line presentation).

The integral on the r.h.s. of Eq. (\ref{partition-func}) which gives
rise to
single-particle partition function $z_M(\beta)$ is
defined in the rapidity range $[-Y_{\rm cm},Y_{\rm cm}]$, where
$Y_{\rm cm}=Y_{\rm beam}/2$.
Then, the functions $\varphi_M(y)$
are normalized to unity in the same range.
If one integrates Eq. (\ref{y-fit}) with respect to rapidity in
this range
it is easy to find that a result of integration on the r.h.s. equals
to the sum of all coefficients $C(M)$ plus $C_{\rm therm}$.
At the same time the value of this integral equals the
area under the ``rapidity'' curve (solid, thick blue curve) in
Fig.\ref{figure:NA49-Mmax12} or to the total number of participated
protons which would be
registered in case if the total rapidity window
$[-Y_{\rm cm},Y_{\rm cm}]$ is allowed experimentally:
$N_p^{\rm (tot)}=\sum_M C(M)+C_{\rm therm}$.
For instance, for NA49 experimental data
\cite{NA49-PRL-v82p2471-1999}, we obtain $N_p^{\rm (tot)}=151$.

So, every coefficient $C(M)$ tells us how many protons undergo $M$
effective
collisions or what is the popularity of every collision ensemble.
For instance, ensemble of the protons which participated
in nine effective collisions, $M=9$, consists of
$10$ protons, i.e. $C(9) \simeq 10$.
What is very important, we learn from
this expansion that $C_{\rm therm}\simeq 18$, it means that approximately
eighteen protons come from a thermal source what makes up $12$\%
(twelve percent) of all participated protons.

Now we would like to draw attention to the ensemble with maximal
collisions, $M=12$.
This value was determined from UrQMD \cite{urqmd} evaluation  of the
mean maximum number of the effective nucleon collisions.
It turns out, as it is seen from
Fig.\ref{figure:NA49-Mmax12} (upper panel), that the partial function
$\varphi_{12}(y)$ does not ``fill in'' successfully
the central rapidity region.
That is why the presence of the thermal function
$\varphi_{\rm therm}(y)$ (Gaussian like curve in the
center of Fig.\ref{figure:NA49-Mmax12}) is necessary in the expansion
(\ref{y-fit}).
Even this big number of effective collisions, $M_{\rm max}=12$, cannot
give rise to a source which is compared in the central rapidity region
with a thermal one.
It is the main reason why the thermal source should be presented in
the expansion (\ref{y-fit}).

Note, in the case of finite and small number of experimental
points the set of functions, $\varphi_M(y)$, is overcomplete.
To choose a unique configuration of the variable parameters we
use the maximum entropy method \cite{soroko}.

In this analysis we are coming to one of the main conclusions,
which can be derived from our model:
The thermal source has absolutely different nature of origination, it
cannot be created just due to the hadron reactions of nucleons which
result in randomization and subsequent isotropization of the nucleon
momentum.
The thermal source can emerge as a result of appearance ``at once'' of
many new degrees of freedom.
We know just one candidate to this role, it is the quark-gluon plasma,
for instance, its creation can occur in collision of
nucleons in the presence of a dense medium,
$N+N \to n_g + n_q$.
Then, a many-parton system, which emerges in the collision,
consists of $n_g \gg 1$ gluons and $n_q \gg 1$ quarks.
All momenta of quarks and gluons can be regarded from the very
beginning as random ones
and thermalization of the system occur during a time span
$\tau_{\rm therm}=0.6$~fm/c \cite{Gyulassy-0709.171}.
Hence, the protons which come from the thermal source indicate
the presence of the QGP in the fireball and we can determine a
power of the QGP by the number of protons outcoming from the thermal
source or by the value of $C_{\rm therm}$.

Actually, the total number of nucleons which appear as a result of
hadronization of the QGP can be then evaluated with accounting for
isotope composition of the colliding nuclei:
$\dis N_{\scriptstyle N}^{\rm (QGP)}=C_{\rm therm} \, \frac{A}{Z}$.
For instance, in the experiment under consideration
we find $C_{\rm therm}\simeq 18$, and then
$N_{\scriptstyle N}^{\rm (QGP)} \simeq 46$, i.e. approximately $46$
nucleons were created by the QGP or by several QGP drops.
This makes up $12\%$ from a total number of net nucleons which are
the participants of the collision,
$N_{\scriptstyle N}^{\rm (participants)} \simeq 382$.
Then, we estimate a ``nucleon power'' of the QGP,
$P_{\rm qgp}\equiv N_{\scriptstyle N}^{\rm (QGP)}
/N_{\scriptstyle N}^{\rm (participants)}$,
which was created in nucleus-nucleus collision.
For instance, it turns out that
$P_{\rm qgp}\simeq 12\%$ in Pb+Pb collisions (SPS) at
$E_{\rm kin}=158$A GeV.

The same analysis was carried out for proton distribution from
11.6A GeV/c Au + Au collisions measured by the E802 Collaboration
\cite{E802-PRC-v60-064901-1999}.
The fit to the experimental data (0-3\% centrality) allows to extract
the values of parameters which are shown in Table 2.
For this experiment  UrQMD \cite{urqmd}
evaluation gives  the mean maximum number of the
effective nucleon collisions $M_{\rm max} = 13$.
The width of the  rapidity window in this
experiment avoids the necessity to take into account the collision
ensemble $M=1$ too.
\noindent\begin{tabular}{c c c c c c c c }
&&&&&&& Table 2 \\
\hline%
   \multicolumn{1}{|c}{$C(2)$} & \multicolumn{1}{|c}{$C(3)$}
 & \multicolumn{1}{|c}{$C(4)$} & \multicolumn{1}{|c}{$C(5)$}
 & \multicolumn{1}{|c}{$C(6)$} & \multicolumn{1}{|c}{$C(7)$}
 & \multicolumn{1}{|c}{$C(8)$} & \multicolumn{1}{|c|}{$C(9)$} \\%
\hline%
   \multicolumn{1}{|c}{12.2} & \multicolumn{1}{|c}{14.5}
 & \multicolumn{1}{|c} {6.5} & \multicolumn{1}{|c} {5.1}
 & \multicolumn{1}{|c} {5.1} & \multicolumn{1}{|c} {5.8}
 & \multicolumn{1}{|c}{6.8}  & \multicolumn{1}{|c|}{8.2} \\%
\hline
\end{tabular}

\noindent\begin{tabular}{ c c c c c c c}
\hline%
   \multicolumn{1}{|c}{$C(10)$}
 & \multicolumn{1}{|c}{$C(11)$} & \multicolumn{1}{|c}{$C(12)$}
 & \multicolumn{1}{|c}{$C(13)$} & \multicolumn{1}{|c}{$C_{\rm therm}$}
 & \multicolumn{1}{|c|}{$p_{\rm max}$ (GeV/c)} \\%
\hline%
   \multicolumn{1}{|c}{9.7}
 & \multicolumn{1}{|c}{11.4} & \multicolumn{1}{|c}{13.0}
 & \multicolumn{1}{|c}{14.7} & \multicolumn{1}{|c}{37.9}
 & \multicolumn{1}{|c|}{0.724} \\%
\hline
\end{tabular}

\medskip

\begin{figure}
\includegraphics[width=0.47\textwidth] {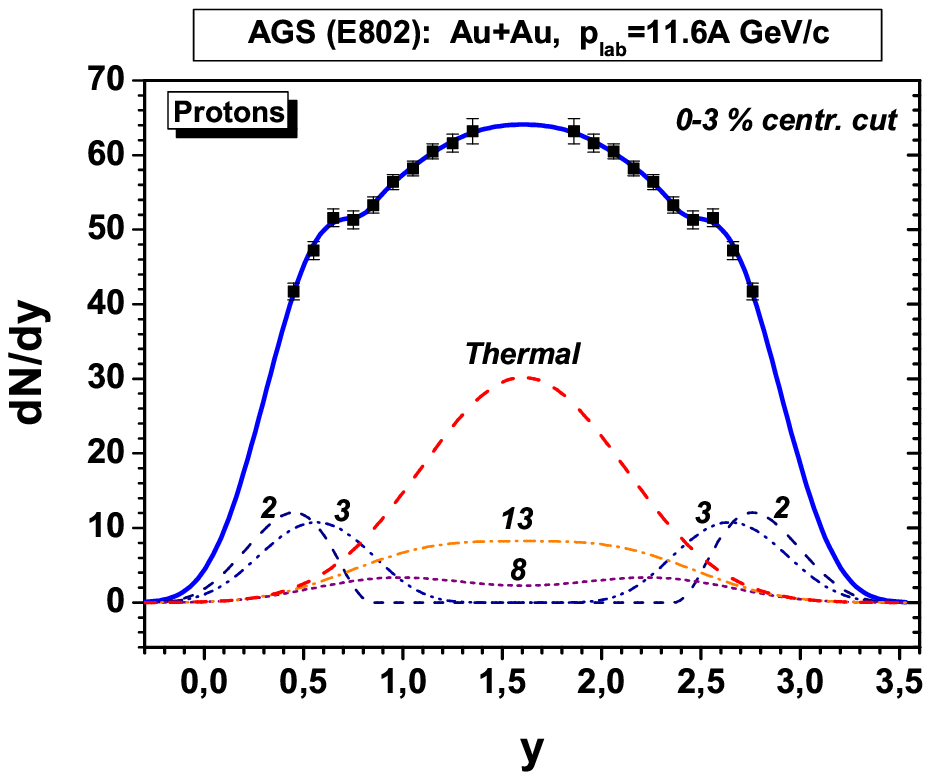}
\includegraphics[width=0.47\textwidth] {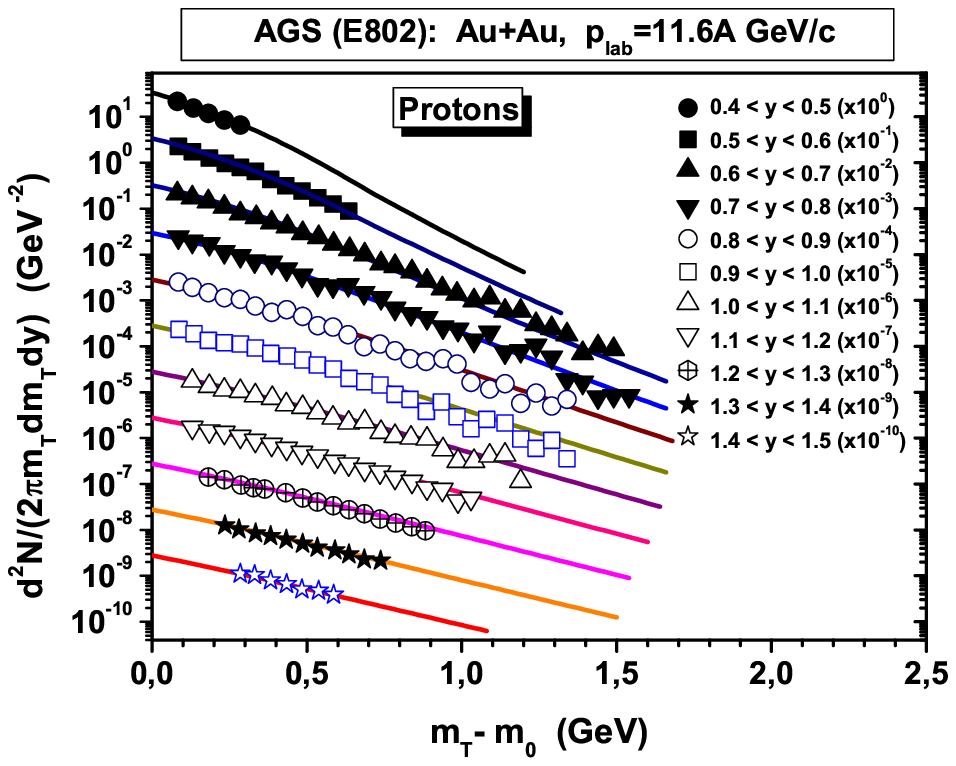}
\caption{\noindent
{\it Upper panel:} The result of the fit (thick solid
curve, blue in on-line presentation) to experimental
  data \cite{E802-PRC-v60-064901-1999} on rapidity distribution
  (see Eq. (\ref{y-fit})).
  Broken curves marked by the numbers $M$ represent the partial
  contributions from every collision ensemble,
  $C(M)\cdot \varphi_M(y)$, the thermal contribution is represented by
  central Gaussian-like dashed curve (red in on-line presentation).\\
  {\it Lower panel:}  The solid curves  represent the evaluation of the
  $m_{\perp}$-spectra   obtained in accordance
  with Eq. (\ref{tr-total})  were we use the same values of the
  coefficients $C(M)$   which were obtained as a result of the fit of
  the $dN/dy$ data.
  Experimental data points are from
  \cite{E802-PRC-v60-064901-1999}. }
\label{figure:E802-rapidity}
\end{figure}

\noindent
Theoretical curves and experimental data  are depicted in Fig.
\ref{figure:E802-rapidity}.
Notations and marks have the same meaning as in the previous figure.
In the lower panel the solid curves  represent the evaluation of the
$m_\bot$-spectra in different rapidity  windows.
Remind, these curves are obtained without additional fitting of
the data.
We just use the values of parameters from the Table 2.
Meanwhile, on the first step, before fitting the rapidity
distribution, we estimate with the help of the thermal
distribution the slope parameter $\beta$ and find $T = 280$~MeV.

We can estimate as well the nucleon power of the produced QGP in experiment
Au~+~Au at 11.6A ~GeV/c (0-3\% centrality).
In this case $C_{\rm therm} \simeq 38$ and $N_p^{\rm (tot)} \simeq 151$,
hence we obtain $P_{\rm qgp} \simeq 25\%$.

{\bf Summary and discussion. }
In the proposed Maximum Isotropization Model
\cite{anchishkin-ezhov-ujp2007} the maximum number of collisions
(reactions), $M_{\rm max}$, assumed to be finite and determined by the
nuclear number $A$, initial energy and centrality.
With the help of
the UrQMD transport model \cite{urqmd} it was found that for SPS
(Pb+Pb, 158A GeV) conditions \cite{NA49-PRL-v82p2471-1999}, $M_{\rm
max}=12$, and for AGS (Au+Au, 11.6A GeV/c, 0-3\% centrality)
conditions \cite{E802-PRC-v60-064901-1999}, $M_{\rm max}=13$.
Utilizing thermal distribution we extract a slope parameter from
experimental data on the proton $m_{\perp}$-spectra: for SPS
conditions \cite{NA49-PRL-v82p2471-1999}, $T=248$~MeV, and for AGS
conditions \cite{E802-PRC-v60-064901-1999} (0-3\% centrality),
$T=280$~MeV.
It is evidently seen from Fig.~\ref{figure:NA49-Mmax12}
(lower panel) that the $m_{\perp}$-spectrum is mainly determined by
the thermal component, and in any case the slope of the partial
contribution, marked by $M$, is approximately the same as of thermal
distribution.
Exactly of that reason the $m_{\perp}$-spectrum is low
informative about collision ensembles or the information about
rescattered nucleons almost lost in this presentation. On the other
hand, the $m_{\perp}$-spectrum as a trigger gives possibility to
extract a value of the slope parameter.

Next, we made the fit of experimental data
\cite{NA49-PRL-v82p2471-1999,E802-PRC-v60-064901-1999} on the
rapidity distribution of the net protons and obtained the set of
coefficients $C(M)$ (see Tables~1,~2) which are nothing more as an
absolute number of protons in every collision ensemble. Note, the
proton data is interesting first of all because we know an exact
value of the initial nucleon momentum. As a matter of fact, the
partial expansion, $dN/dy=\sum_{M=M_{\rm min}}^{M_{\rm max}}C(M) \,
\varphi_M(y)$ (see (\ref{y-fit})), where we use obtained
coefficients $C(M)$ from Tables~1,~2, makes a good description of
the experimental data on rapidity distribution, except the central
rapidity region. It means that the central rapidity region cannot be
described just by finite number of nucleon rescatterings (hadron
reactions).
Then, we are forced to take into account also the
thermal source, which evidently has a different nature. We assume
that this source is a thermalized multi-parton system (QGP)
\cite{Gyulassy-0709.171} which through hadronization process emits
totally thermalized nucleons.
The knowledge of the number of
protons, $C_{\rm therm}$, which come from the QGP, gives us a
possibility to evaluate the ``nucleon power'' of the QGP, $P_{\rm qgp}$,
created in the particular experiment on nucleus-nucleus collision.
We find that for SPS conditions \cite{NA49-PRL-v82p2471-1999},
$P_{\rm qgp} \simeq 12\%$, and for AGS conditions
\cite{E802-PRC-v60-064901-1999} (0-3\% centrality), $P_{\rm
qgp}\approx 25\%$.
So, following the proposed criterium we can claim that QGP
(as a nucleon source)  was created not only at SPS energies
\cite{heinz2000} but it was also created, even more powerful with
respect to nucleons, in the central collisions at AGS energies.
Meanwhile, in accordance with
UrQMD estimations the number of pions created in hadron reactions at
the SPS \cite{NA49-PRL-v82p2471-1999} and AGS
\cite{E802-PRC-v60-064901-1999} energies are approximately the same.
Hence, the number of pions created by the thermal source at the SPS
is much bigger than the number of pions created by the thermal
source at the AGS.
From that we can conclude that ``pion power'' of
QGP created in nucleus-nucleus collisions at the SPS up to one order
higher than that one created at the AGS.

All this leaves us with the continued challenge of applying the
model to other experiments and problems.


{\bf Acknowledgements:}
Authors would like to express their gratitude to A.~Muskeyev for
providing them with results of UrQMD calculations.
D.A. thanks E.~Martynov for useful instructions of handling of
MINUIT.
S.Ye. is thankful to J.-P.~Blaizot for support and warm hospitality
during his visit to the ECTP (Trento, Italy).

\end{document}